\documentclass[twocolumn,aps]{revtex4}
\usepackage{amsmath}
\usepackage{bm}
\usepackage{graphicx}
\usepackage{mathrsfs}
\usepackage{footmisc}
\usepackage{multirow}
\usepackage{graphicx}
\usepackage{booktabs, ctable}
\usepackage{xcolor, color, framed}

\begin{document}

\title{ $\psi'$ photoproduction in high energy nuclear collisions}
\author{Baoyi Chen$^{1,2}$, Carsten Greiner$^2$, Wei Shi$^1$, Wangmei Zha$^3$, and Pengfei Zhuang$^4$}
\affiliation{$^1$Department of Physics, Tianjin University, Tianjin 300350, China\\
             $^2$Institut f\"ur Theoretische Physik, Goethe-Universit\"at Frankfurt, D-60438 Frankfurt am Main, Germany\\
             $^3$Department of Modern Physics, University of Science and Technology of China, Hefei 230026, China\\
             $^4$Physics Department, Tsinghua University and Collaborative Innovation Center of Quantum Matter, Beijing 100084, China}
\pacs{25.75.-q, 12.38.Mh, 25.75.Ld, 24.10.Nz}

\date{\today}

\begin{abstract}
We study coherent photoproduction of $\psi'$ and the subsequent suppression in high energy nuclear collisions. The hot medium effect on charmonium photoproduction is controlled by the competition between the fireball size and nuclear size. The fireball size is 
related to the hadronic collisions in the overlap area of two colliding nuclei, while coherent photoproduction is distributed over the target nucleus. When the former is smaller than the later, the photoproduction is only slightly affected by the hot medium. We calculate the yield ratio of $\psi'$ to $J/\psi$ in semi-central Pb-Pb collisions at LHC energy, while the high $p_T$ limit is characterized by 
the fireball and the hadroproduction, its value in low $p_T$ limit can still reach the significant enhancement by photoproduction.
\end{abstract}
\maketitle

\section{Introduction}
Charmonia are considered as a sensitive signal of the quark-gluon plasma (QGP) formation in high energy nuclear collisions~\cite{Matsui:1986dk}. Due to the different binding energies for the ground ($J/\psi$) and excited ($\psi', \chi_c,......$) states, the initially produced charmonia are sequentially dissociated in hot medium~\cite{Satz:2005hx} and their suppressions in the final state are used to measure the fireball temperature. Including the cold (such as shadowing effect~\cite{Mueller:1985wy}, Cronin effect~\cite{Cronin:1974zm} and nuclear absorption~\cite{Gerschel:1988wn}) and hot (for instance color screening~\cite{Satz:2005hx} and regeneration~\cite{BraunMunzinger:2000px,Thews:2000rj,Yao:2017fuc}) nuclear matter effects before and after the hot medium formation, charmonia in heavy ion collisions are deeply investigated at RHIC~\cite{Grandchamp:2002wp,Liu:2009nb,Zhao:2010nk} and LHC~\cite{Zhou:2014kka,Chen:2015iga,Chen:2012gg,Zhao:2011cv} energies.

In addition to the charmonium hadroproduction through strong interaction, the charmonium photoproduction through electromagnetic interaction should be taken into account too in high energy nuclear collisions, due to the extremely strong electromagnetic fields created in these collisions. For instance, the magnetic field strength can reach $|eB|\simeq 10\ m_\pi^2$ at LHC energy~\cite{Deng:2012pc}, where $m_\pi$ is the pion mass in vacuum. The fields generated by one nucleus can be approximated as quasi-real photons moving longitudinally (Weizs\"acker-Williams-Method~\cite{vonWeizsacker:1934nji}) and fluctuate into vector hadrons by two gluon exchange with the other nucleus in the lowest order (photon-Pomeron-fusion~\cite{Klein:1999qj}). This coherent electromagnetic interaction opens a new mechanism for the charmonium production in heavy ion collisions. The very large nuclear modification factor $R_{AA}$~\cite{Adam:2015gba} for $J/\psi$ measured at extremely low transverse momentum $p_T$ in ultra peripheral nuclear collisions is far beyond the expectation of hadroproduction but supports the photoproduction mechanism~\cite{Abelev:2012ba,Abbas:2013oua,Bertulani:2005ru,Xie:2015gdj}.

The photoproduction is usually taken into account in ultra peripheral collisions where the hadroproduction is absent~\cite{Xie:cite2,Yu:2014eoa,Yu:2015kva}. When the hadroproduction happens in collisions with nonzero participant nucleons, the photoproduction is expected to be neglected. However, the excess of $J/\psi$ at very low $p_T$ is recently observed also in semi-central collisions~\cite{Adam:2015gba}. Considering the fireball formation in semi-central and central collisions, the finally observed charmonia are controlled by not only the production mechanism but also the suppression mechanism. The hadroproduction, including initial production and regeneration, happens mainly in the overlapped region of the nuclear collision, namely in the formed fireball, and therefore is strongly affected by the hot medium. The photoproduction is, however, a coherent electromagnetic process, and happens mainly on the surface of the colliding nuclei where the temperature is lower and the hot medium effect is weaker. Therefore, the hadroproduction is largely suppressed by the hot medium, but the photoproduction may survive from the fireball. As a result, the photoproduction still dominates the very low $p_T\ J/\psi$s in semi-central collisions~\cite{Shi:2017qep}. Of course, for central collisions where the electromagnetic fields become weak, the hadroproduction is the dominant source of charmonia at any $p_T$.

What is the hot medium effect on $\psi'$ photoproduction? Considering the weaker binding for the excited charmonium states, the $\psi'$ dissociation temperature $T_{\psi'}\simeq T_c$, where $T_c$ is the critical temperature of deconfinement phase transition, is much lower than $T_{J/\psi}\simeq 1.5 T_c$~\cite{Satz:2005hx}, the produced $\psi'$s via photoproduction around the fireball surface may still be eaten up by the hot medium and cannot be seen in the final state. In this paper, we study $\psi'$ photoproduction in semi-central nuclear collisions at LHC energy. Taken into account  the fact that both the ground and excited states suffer from the similar cold medium effect before their formation, we do not calculate the $\psi'$ distribution itself but focus on the ratio of $\psi'$ to $J/\psi$, to eliminate the cold medium effect and focus on the hot medium effect.

\section{Hadroproduction and Photoproduction}

We first consider the charmonium hadroproduction. 
Charmonium production from parton hard scatterings 
in nucleon-nucleon (pp) collisions is a non-perturbative process. The 
hadronization of $c\bar c$ pair into color singlet 
charmonium state $\Psi$ have been studied 
extensively based on NRQCD model~\cite{Ma:2013yla}, 
color singlet model~\cite{Braaten:1996pv}, color evaporation model~\cite{Amundson:1996qr}, 
etc. In this work, 
instead of starting from the partonic process, we 
parametrize the charmonium distribution 
in pp collisions as an input of charmonium evolutions in the nucleus-nucleus (AA) 
collisions, and 
focus on the hot medium effects on different quarkonium bound states. 
The evolution of the charmonia via hadroproduction in the deconfined medium 
is controlled by the Boltzmann transport equation~\cite{Liu:2009nb,Zhou:2014kka,Chen:2015iga},
\begin{equation}
\label{eq-trans}
{\partial f_\Psi\over \partial t} +{\bf v}\cdot \bigtriangledown f_\Psi =-\alpha_\Psi
f_\Psi +\beta_\Psi,
\end{equation}
where $f_\Psi(x,{\bf p}|{\bf b})$ ($\Psi=J/\psi, \psi', \chi_c$) is the charmonium phase space distribution and ${\bf v}={\bf p}/E_\Psi$ with charmonium energy $E_\Psi=\sqrt{{\bf p}^2+m_\Psi^2}$ the charmonium velocity. The second term on the left hand side represents the free streaming process. The hot medium induced suppression and regeneration are respectively controlled by the loss term $\alpha_\Psi$ and gain term $\beta_\Psi$. The former in QGP is dominated by gluon dissociation~\cite{Bhanot:1979vb} $gJ/\psi\to c\bar c$ at finite temperature~\cite{Zhou:2014kka}, and the latter is determined by the inverse process of the gluon dissociation by using the detailed balance between the suppression and regeneration. 

The decay rate of charmonia in QGP is therefore 
connected with parton-charmonium inelastic cross 
section $\sigma_{g\Psi}$~\cite{Zhu:2004nw} and also parton density $f_g$, 
\begin{align}
\label{Cdfactor}
\alpha_\Psi ={1\over 2E_T} \int {d^3{\bf k}\over {(2\pi)^32E_g}}\sigma_{g\Psi}({\bf p},{\bf k},T)4F_{g\Psi}({\bf p},{\bf k})f_g({\bf k},T)
\end{align}
and $F_{g\Psi}$ is the flux factor. $E_g$ and $E_T=\sqrt{m_\Psi^2+p_T^2}$ are the gluon 
energy and charmonium transverse energy, respectively.  
For different bound states, their inelastic cross sections with partons are different, which 
brings different magnitude of decay rates for ($J/\psi$, $\chi_c$, $\psi^\prime$, etc). 

The transport equation can be solved analytically, and the solution in central rapidity region is
\begin{eqnarray}
\label{eq-analy}
&&  f_\Psi({\bf x}_t,{\bf p}_t, \tau|{\bf b})\\
&=& f_\Psi(\tilde{\bf x}_0,{\bf p}_t,\tau_0|{\bf b})e^{-\int_{\tau_0}^{\tau} d\tau_1\alpha_\Psi(\tilde{\bf x}_1,{\bf p}_t,\tau_1|{\bf b})} \nonumber \\
&&+ \int_{\tau_0}^\tau d\tau_1\beta_\Psi(\tilde{\bf x}_1,{\bf p}_t,\tau_1|{\bf b})e^{-\int_{\tau_1}^\tau d\tau_2\alpha_\Psi(\tilde{\bf x}_2,{\bf p}_t,\tau_2|{\bf b})}\nonumber
\end{eqnarray}
with the velocity-dependent coordinate $\tilde {\bf x}_n={\bf x}_t-{\bf v}_t(\tau-\tau_n)$ which indicates the leakage effect in the transverse direction, namely that high speed charmonia can escape from the fireball easily. The first and second terms of the solution come respectively from the initial production at time $\tau_0$ (parametrized with charmonium distribution in pp collisions) and the regeneration at time $\tau_0<\tau_1<\tau$ in the hot medium, both suffer from the suppression in the QGP described by the dissociation rate. The initial distribution is a superposition of nucleon-nucleon collisions, controlled by the colliding energy and nuclear geometry, plus an additional modification by the cold nuclear matter effect. The detailed treatment at LHC energy can be seen in Refs.\cite{Zhou:2014kka,Chen:2016dke}.

Both the dissociation rate $\alpha$ and regeneration rate $\beta$ are temperature dependent. The total dissociation cross section is proportional to the charmonium averaged radius~\cite{Chen:2016dke} which is determined by the potential model with lattice simulated heavy quark potential~\cite{Satz:2005hx}. The regeneration rate via coalescence mechanism~\cite{Fries:2008hs,Hwa:2002tu} is proportional to the heavy quark and antiquark distributions $f_c$ and $f_{\bar c}$~\cite{Zhao:2017yan} which are normally taken as kinetically thermalized distributions~\cite{ALICE:2012ab} at LHC energy. The local temperature and fluid velocity of the hot medium, appearing in $\alpha, \beta, f_c$ and $f_{\bar c}$, are governed by ideal hydrodynamics~\cite{Zhou:2014kka}.

We now discuss the starting time $\tau_0$ of charmonium dissociation in the QGP. It depends on the QGP formation time and the charmonium formation time, and it takes the longer one of the two. For the QGP formed at LHC energy, its formation time is usually taken as $\sim 0.6$ fm/c~\cite{Shen:2011eg}. The $c\bar c$ pairs in heavy ion collisions are initially created by parton fusions $gg (q\bar q)\to c\bar c$ with a small distance $r_{c\bar c}\sim 1/(2m_c)$. To evolve from a pair of $c \bar c$ into a charmonium eigenstate needs about 0.5 fm/c for $J/\psi$ and 1 fm/c for $\psi'$ and $\chi_c$~\cite{Kharzeev:1999bh}. In comparison with the QGP formation time, we assume that $J/\psi$s form before the QGP formation and $\psi'$s after the QGP formation. Therefore, we take the starting time of the hot medium effect on charmonia in (\ref{eq-analy}) as 0.6 fm/c for $J/\psi$ and 1 fm/c for $\psi'$ and $\chi_c$. The different formation time for different charmonium state will not largely affect the charmonium distributions at low $p_T$ where the charmonium production is governed by the regeneration and photoproduction but sizeably change the high $p_T$ distribution where the initial production becomes the dominant \emph{}mechanism.

With Cooper-Frye formula~\cite{Cooper:1974mv}, the number of $\Psi$s via hadroproduction at proper time $\tau$ is
\begin{equation}
N_\Psi^h(\tau|{\bf b}) = {1\over (2\pi)^3}\int d^2{\bf x}_td^2{\bf p}_t\tau m_t f_\Psi({\bf x}_t,{\bf p}_t,\tau|{\bf b})
\end{equation}
with charmonium transverse energy $m_t=\sqrt{m_\Psi^2+p_T^2}$.

\begin{figure}[!htb]
{\includegraphics[width=0.245\textwidth]{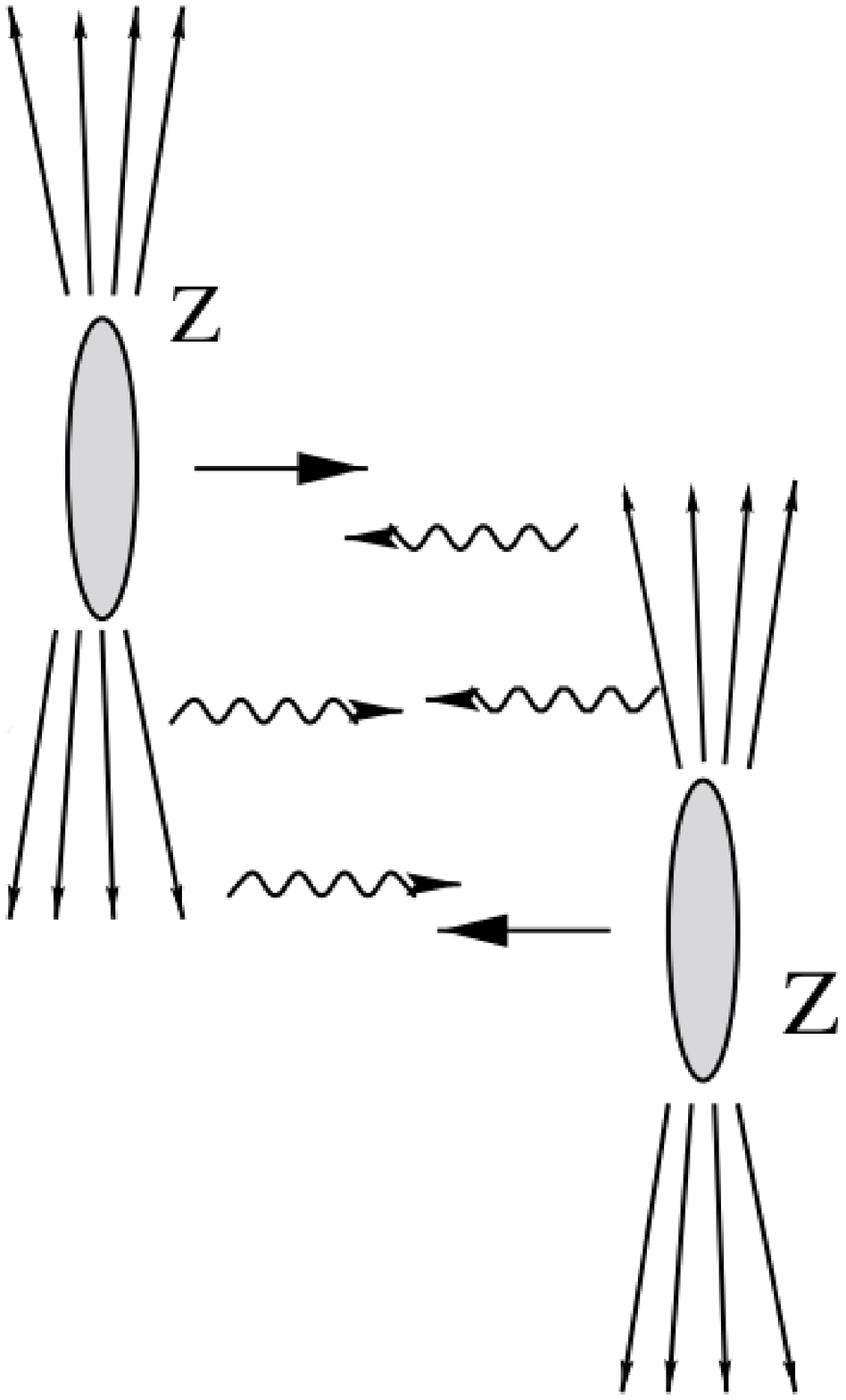}
\includegraphics[width=0.27\textwidth]{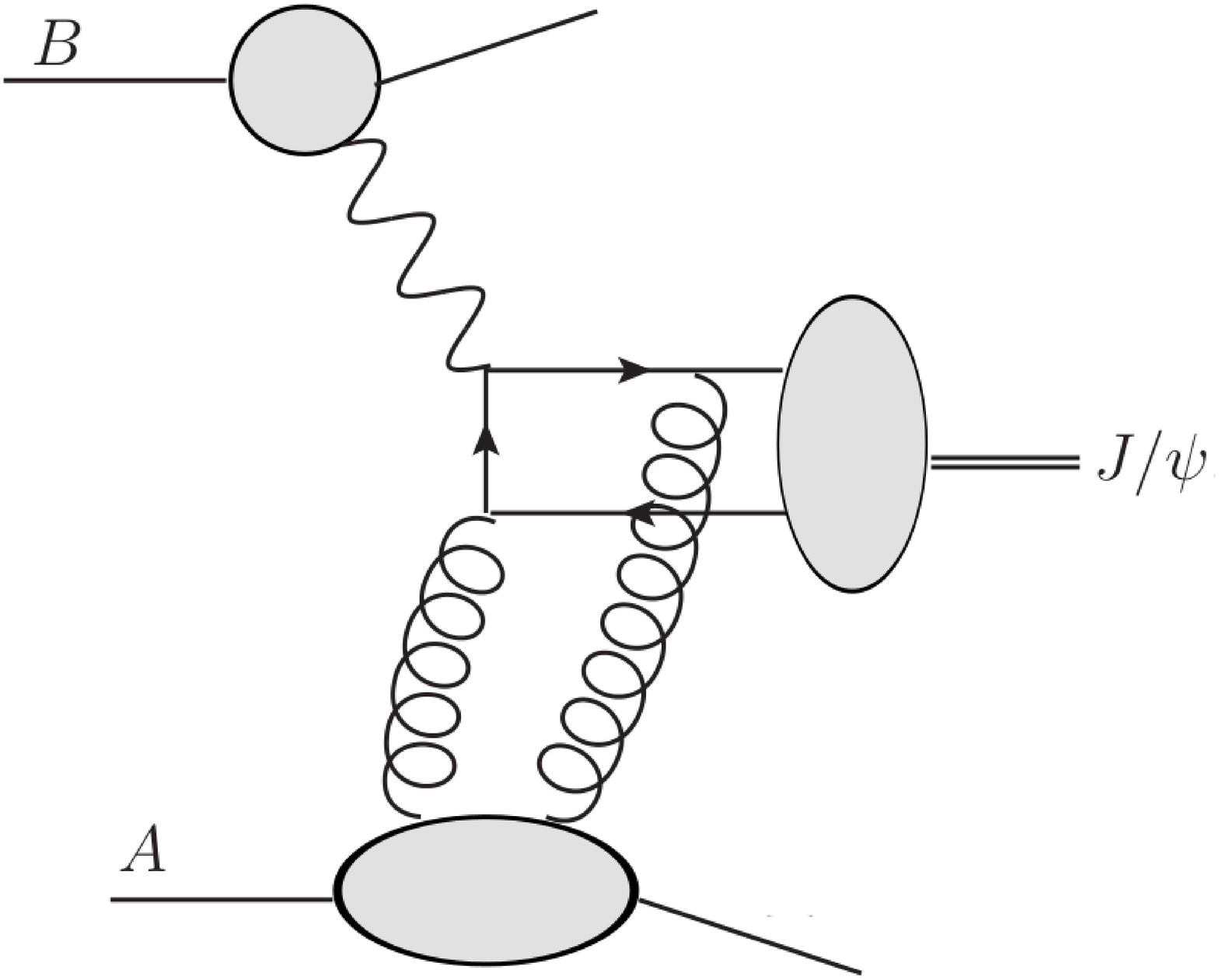}
\caption{Upper schematic figure is to approximate the Lorentz-contracted 
electromagnetic fields as longitudinally 
moving photons~\cite{G:Baur2002}. Lower schematic figure is for 
the coherent photoproduction of color singlet vector mesons (charmonium in this case) from 
photon-nucleus interactions with two gluon exchange~\cite{V:Guzey}.  
}
\label{fig-schematic}}
\end{figure}

We now turn to the evolution of the charmonia via photoproduction. With large magnitudes of electromagnetic fields, the photoproduction can enhance or even dominate the charmonium yield at extremely low $p_T$ in nucleus-nucleus collisions with $b<2R_A$~\cite{Adam:2015gba,Zha:2017etq} where $R_A$ is the colliding nuclear radius. By treating the energy flux of the electromagnetic fields created by one colliding nucleus as equivalent quasi-real photons, one can approximate the vector meson production through the electromagnetic interaction with the other nucleus to be a scattering process between photons and the nucleus, $\gamma A\to \Psi A$, see Fig.\ref{fig-schematic}. 
The photon energy spectrum is determined by the Poynting vector of the electromagnetic fields 
$\int_{-\infty}^{\infty} d\tau\int d{\bf x}_t\cdot({\bf E}_t
\times {\bf B}_t) =\int_0^\infty dw w{dN_\gamma\over dw}$~\cite{Vidovic:1992ik} with 
the conservation of 
the energy flux through the transverse plane, 
\begin{eqnarray}
\label{eq-spectrum}
&&{dN_\gamma\over dw}(w|{\bf b}) \nonumber \\
&=& {1\over \pi w} \Big|{\bf E}_t(w|{\bf b})\Big|^2\nonumber\\
&=& {Z^2\alpha\over 4\pi^3 w}\Big|\int_0^\infty dk_t k_t^2{F(k_t^2 +w^2/\gamma_l^2)\over k_t^2 +w^2/\gamma_l^2}J_1(k_tb)\Big|^2,
\end{eqnarray}
where $Ze$ is the nuclear charge, $\gamma_l=\sqrt{s_{NN}}/(2m_N)$ the nucleon Lorentz factor, $k_t$ the photon transverse momentum, $J_1$ the first kind Bessel function, and $\alpha=e^2/(\hbar c)$ the electromagnetic coupling
constant. The nuclear form factor $F(q^2)$ is the Fourier transform of the charge distribution in the nucleus. 
The photon-nucleus coherent cross section $\sigma_{\gamma A}$ 
can be scaled from photon-proton scattering cross section, 
\begin{align}
\label{eq-gammaA}
&\sigma_{\gamma A\rightarrow J/\psi A} \nonumber \\ 
&= {\alpha \sigma_{tot}^2(J/\psi A)\over 4f_V^2} 
\int_{-t_{min}}^\infty dt |F(t)|^2  \nonumber \\  
&= {\alpha \over 4f_V^2} [\int d^2{\bf x}_t (1-e^{-\sigma_{tot}(J/\psi p)T_A(
{\bf x}_t)})]^2
\int_{-t_{min}}^\infty dt |F(t)|^2 ,
\end{align}
$-t_{min}=[M_{J/\psi}^2/(4w\gamma_L)]^2$ is 
the minimum momentum transfer 
squared to produce a vector meson with mass $M_{J/\psi}$ in the 
laboratory reference frame. $f_V^2/4\pi =10.4$ is fitted with $J/\psi$ mass 
and its leptonic decay width~\cite{Klein:1999qj}. 
$T_A$ is the thickness function of target nucleus.  
$J/\psi$-proton total scattering cross section is obtained from~\cite{Klein:1999qj, 
Shi:2017qep,book:HERA} 
fitted with HERA experimental data. 

The strong electromagnetic fields can not only produce vector mesons and dileptons, but also 
break the target nucleus, which will change the nuclear charge distributions and therefore 
the form factors in the formula of photon density Eq.(\ref{eq-spectrum}) 
and also photon-nucleus cross section Eq.(\ref{eq-gammaA}). This mechanism becomes the main 
source for the nuclear breakup in \emph{Ultra-peripheral} collisions with $b>2R_A$. However, 
this effect is much smaller compared with the contributions of 
hadronic collisions at $b<2R_A$, and is neglected 
in this work.

With both the photon density and $\gamma A$ cross section, 
one can 
obtain the charmonium number via photoproduction (For more details about charmonium final
photoproduction with hot medium effect, please see \cite{Shi:2017qep}),
\begin{eqnarray}
\label{eq-photo}
N_\Psi^p(\tau|{\bf b}) &=& \int d^2{\bf x}_t\tilde f_\Psi(w,{\bf x}_t+{{\bf b}\over 2})e^{-\int_{\tau_0}^\tau d\tau_1\alpha_\Psi({\bf x}_t,{\bf p}_t,\tau_1|{\bf b})}\nonumber\\
&& +(y\rightarrow -y, {{\bf b}\over 2}\rightarrow -{{\bf b}\over 2}),
\end{eqnarray}
where the charmonium rapidity $y$ in laboratory frame is related to the photon energy $y=\ln[2w/M_\Psi]$, $\tilde f_\Psi$ is the produced $\Psi$ number per unit of transverse area through the process $\gamma A\rightarrow \Psi A$, including the modification from the cold nuclear matter effects~\cite{Shi:2017qep}, and the exponential factor indicates the suppression by the hot medium effect. Note that quasi-real photons interact with the entire nucleus coherently, and the produced charmonia are over the entire nuclear surface. This is very different from the hadroproduction where charmonia are produced mainly in the overlap region of the two colliding nuclei where the QGP is formed. With the different spatial distributions, the produced charmonia via hadroproduction and photoproduction will suffer from different suppression, and this will help us to distinguish the two production mechanisms.

The finally observed charmonia number is $N_\Psi^h + N_\Psi^p$ at proper time $\tau\to\infty$. Considering the decay contribution from excited states to ground state, we take for $J/\psi$ in the final state $60\%$ from the direct production (including both initial production and regeneration) and $40\%$ from $\psi'$ and $\chi_c$ decay.

\section{Hydrodynamic equations for QGP evolutions} 
The deconfined medium with large pressure gradient in the transverse plane expands outside 
violently after the nuclear collisions, with a lifetime of a few $fm/c$.  
One can employ the hydrodynamic equations for the QGP 
spatial and time evolutions, 
\begin{align}
\label{eq-hydro}
\partial_\mu [(e+p)u^\mu u^\nu -g^{\mu\nu}p] =0
\end{align}
here $e$ and $p$ is the local energy density and the pressure of QGP. $u^\mu$ is the four 
velocity of the fluid cells in the lab rest frame. 
With the assumption of Hubble-like longitudinal expansion, hydrodynamic equations are 
simplified as 2-dimensional transverse accelerations and the longitudinal expansion with 
a constant velocity.   
The initial maximum temperature of the QGP in the most central collisions (impact 
parameter $b=0$) is fitted as $T_0^{\tau_0}({\bf x}_T=0)=485$ MeV and 510 MeV for 
$\sqrt{s_{NN}}=2.76$ TeV and $5.02$ TeV Pb-Pb collisions respectively  
from the final charge 
multiplicities. The lifetime of QGP produced 
at the impact parameters of $b=0$ and $b=8.4$ fm 
is around $\sim 10$ fm/c and $\sim 6$ fm/c respectively in these two 
energies. $\tau_0\sim 0.6$ fm/c is the time scale of QGP reaching local equilibrium 
where QGP starts transverse acceleration. Note that QGP local temperature depends on 
both time and spatial coordinates which results in different suppressions on the 
charmonia from coherent photoproduction and hadroproduction.

\section{Numerical Results}
\begin{figure}[!htb]
{\includegraphics[width=0.45\textwidth]{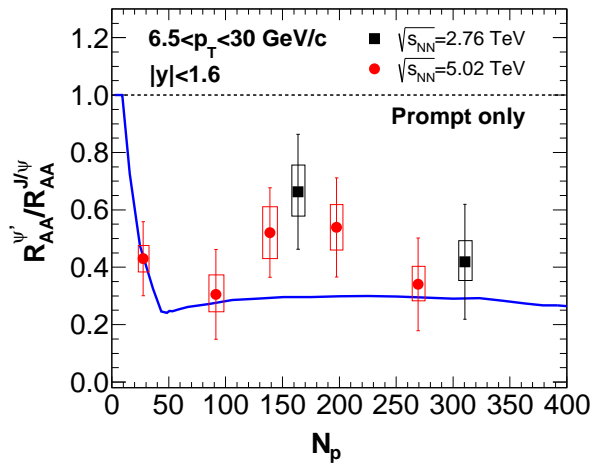}
\caption{(Color online) The double ratio $R_{AA}^{\psi'}/R_{AA}^{J/\psi}$ for prompt charmonia as a function of participant number $N_p$ in Pb-Pb collisions. The solid line is the model calculation with only initial production at colliding energy $\sqrt{s_{NN}}=5.02$ TeV, and the experimental data are from the CMS collaboration~\cite{Khachatryan:2014bva,Scomparin:2017pno, Sirunyan:2016znt}. }
\label{fig1}}
\end{figure}

We first consider the charmonium distribution at extremely high $p_T$ where the regeneration and photoproduction can be safely neglected and the initial production becomes the only charmonium source. To have a direct comparison with the CMS data for prompt charmonia, we do not include the feedback from the $B$-mesons. The model calculation of the double ratio $R_{AA}^{\psi'}/R_{AA}^{J/\psi}$ in Pb-Pb collisions at colliding energy $\sqrt {s_{NN}}=5.02$ TeV as a function of centrality and the comparison with the CMS data in the region $6.5 <p_T< 30$ GeV/c at $\sqrt{s_{NN}}=2.76$ and $5.02$ TeV are shown in Fig.\ref{fig1}. Since the excited state $\psi'$ is loosely bound and the ground state $J/\psi$ is tightly bound, $\psi'$ is much more easily dissociated in the QGP, and the double ratio is always less than unit~\cite{Chen:2013wmr}. With increasing number of participants, the fireball temperature goes up and the suppression for both $J/\psi$ and $\psi'$ increases. In our model calculation, the double ratio in semi-central and central collisions is almost a constant, which is in reasonable agreement with the CMS data at $\sqrt{s_{NN}}=5.02$ TeV. The turning point located at $N_p\simeq 50$ corresponds to the $\psi'$ dissociation temperature.

\begin{figure}[!htb]
{\includegraphics[width=0.45\textwidth]{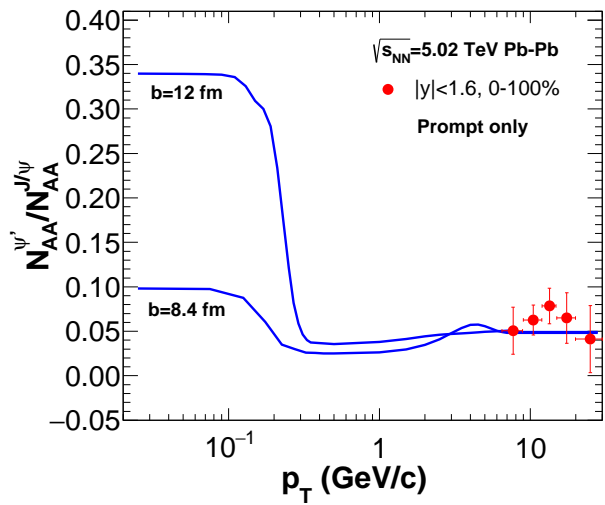}
\caption{(Color online) The yield ratio $N_{AA}^{\psi'}/N_{AA}^{J/\psi}$ for prompt charmonia as a function of transverse momentum $p_T$ in Pb-Pb collisions with colliding energy $\sqrt{s_{NN}}=5.02$ TeV. The two solid lines are the model calculations for events with impact parameter $b=12$ fm and $8.4$ fm (corresponding to $N_p\sim 35$ and $150$ for comparison with Fig.\ref{fig1}), and the data are from the CMS collaboration~\cite{Sirunyan:2016znt} for minimum bias events. }
\label{fig2}}
\end{figure}

In order to see the competition between hadroproduction and photoproduction, we turn to calculate the charmonium $p_T$ distribution in semi-central events. In Fig.\ref{fig2} we show the yield ratio $N_{AA}^{\psi'}/N_{AA}^{J/\psi}$ as a function of $p_T$ for prompt $J/\psi$ and $\psi'$ in Pb-Pb collisions at colliding energy $\sqrt{s_{NN}}=5.02$ TeV. Since the coherent photoproduction $\gamma A \to \Psi A$ is not a superposition of $\gamma p\to \Psi p$ processes, it is better to directly calculate the yield ratio than the $R_{AA}$ ratio. The two solid lines in Fig.\ref{fig2} are the model calculations for two semi-central collisions with impact parameter $b=12$ fm and $8.4$ fm, and the experimental data at high $p_T$ in minimum bias events, corresponding to $b=8.4$ fm, are extracted from the CMS data~\cite{Sirunyan:2016znt} of the $R_{AA}$ ratio, $N_{AA}^{\psi'}/N_{AA}^{J/\psi} = R_{AA}^{\psi'}/R_{AA}^{J/\psi}\times [(d\sigma_{pp}^{\psi'}/dy)/(d\sigma_{pp}^{J/\psi}/dy)]$, where the ratio of $\psi'$ to $J/\psi$ in p-p collisions is $0.166$~\cite{Acharya:2017hjh} without significant $p_T$ dependence.

\begin{figure}[!htb]
{\includegraphics[width=0.45\textwidth]{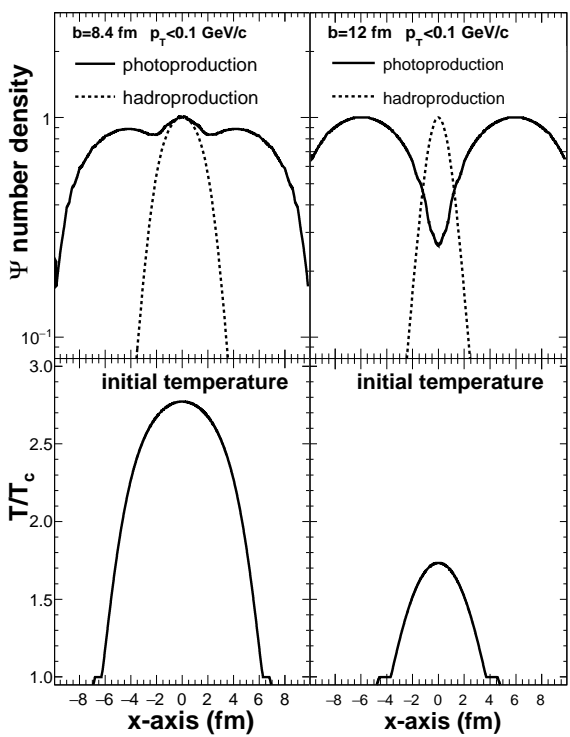}
\caption{ The initial QGP distribution (lower panel) and scaled charmonium number distribution (upper panel) along the $x$ axis in the transverse plane ($x-y$ plane). The impact parameter is chosen as $b=8.4$ fm (left panel) and $12$ fm (right panel). The dashed and solid lines in the upper panel are respectively the scaled charmonium numbers from hadroproduction and photoproduction. }
\label{fig3}}
\end{figure}

In the limit of high $p_T$, there are no photoproduction and regeneration, the charmonium yield is controlled by the initial production and the subsequent suppression in the QGP phase. From the hydrodynamic calculation which determines the local temperature and velocity of the fireball, the initial QGP distribution in the transverse plane is shown in the lower panel of Fig.\ref{fig3} for the two semi-central events. The corresponding space distribution of the $\psi'$ number density through hadroproduction is shown as dashed lines in the upper panel, where the number is scaled by its maximum value. The initial QGP with temperature $T>T_c$ is created in the region of $|x|<4$ fm/c for events with $b=12$ fm and $|x|< 6$ fm/c for events with $b=8.4$ fm. In both cases, the QGP region covers the corresponding hadroproduction region $|x| < 2.5$ fm and $|x| < 4$ fm. Therefore, the strong hot medium effect on $\psi'$ leads to a small yield ratio $N_{AA}^{\psi'}/N_{AA}^{J/\psi}\simeq 0.05$ in both events which is in good agreement with the CMS data~\cite{Sirunyan:2016znt}.

In the other limit of extremely low $p_T$ shown in Fig.\ref{fig2}, the charmonia come mainly from the photoproduction. The scaled $\psi'$ number density distribution from photoproduction is plotted as solid lines in the upper panel of Fig.\ref{fig3}. From the comparison of the photoproduction region with the initial QGP region, most of the produced $\psi'$s are eaten up by the hot medium in event with $b=8.4$ fm, which results in a small yield ratio $N_{AA}^{\psi'}/N_{AA}^{J/\psi}\simeq 0.1$. However, for event with $b = 12$ fm, most of the electromagnetically produced $\psi'$s are distributed outside the QGP fireball and therefore not affected by the hot medium. Since the hadroproduction part is almost fully absorbed by the QGP, the total yield ratio approaches to the coherent photoproduction limit measured in ultra peripheral collisions, $N_{AA}^{\psi'}/N_{AA}^{J/\psi}=0.34$~\cite{Adam:2015sia}.

The observable of $N_{AA}^{\psi\prime}/N_{AA}^{J/\psi}$ in extremely low $p_T$ region 
in semi-central collisions $b<2R_A$ can also help to 
clarify the photoproduction process of vector 
mesons and dileptons. One argument is that particle photoproductions happen 
 before the hadronic collisions, i.e., the electromagnetic fields for 
photoproduction are generated by the entire source nucleus, and interact with the 
entire target nucleus. The other argument is photoproductions happen after the hadronic 
collisions. The electromagnetic fields generated by spectators in the source 
nucleus interact with spectators in the target nucleus. Nuclear electric charges in 
the area of hadronic collisions do not contribute to the photoproduction. In the 
``nucleus-nucleus" scenario which is employed in this work, QGP effects can strongly suppress 
the $\psi\prime$ production, make $N_{AA}^{\psi\prime}/N_{AA}^{J/\psi}$ (b=8.4 fm) 
smaller than the value 
in UPC, see Fig.\ref{fig2}. In ``spectator-spectator" scenario, however,  
both coherently photoproducted $J/\psi$ and $\psi\prime$ are distributed over the area of 
spectators and outside of QGP. This will make $N_{AA}^{\psi\prime}/N_{AA}^{J/\psi}$ in 
semi-central collisions ($b=8.4$ fm) similar to the value in UPC.  
Whether $N_{AA}^{\psi\prime}/N_{AA}^{J/\psi}$ 
are suppressed or not in $p_T<0.1$ GeV/c in semi-central collisions, 
can help to distinguish the two different mechanism of particle photoproduction.

\section{summary}
In summary, we investigated the $\psi'$ photoproduction and the subsequent hot medium effect in high energy nuclear collisions. Since the photoproduction is a coherent process, it takes place on the two colliding nuclear surfaces, and the hot medium effect is weaker in comparison with the hadroproduction. As a result, when the produced fireball is smaller than the colliding nuclear, the charmonia at extremely low $p_T$ is still controlled by the photoproduction. For Pb-Pb collisions with impact parameter $b\sim 12$ fm at LHC energy, while the yield ratio of $\psi'$ to $J/\psi$ is dominated by the hot medium in high $p_T$ limit, its low $p_T$ limit is precisely described by the photoproduction without visible medium effect.

\vspace{0.5cm}
{\bf Acknowledgement:} B.Chen thanks Moritz Greif for helpful discussions. The work is supported by NSFC Grants 11335005, 11705125 and 11775213 and CSC-DAAD Postdoc Scholarship.


\begin{thebibliography}{20}
\bibitem{Matsui:1986dk}
  T.Matsui and H.Satz, Phys. Lett. {\bf B178}, 416(1986).
\bibitem{Satz:2005hx}
  H.~Satz, J. Phys. {\bf G32}, R25(2006).
\bibitem{Mueller:1985wy}
  A.H.Mueller and J.W.Qiu, Nucl. Phys. {\bf B268}, 427(1986).
\bibitem{Cronin:1974zm}
  J.W.Cronin, H.J.Frisch, M.J.Shochet, J.P.Boymond, R.Mermod, P.A.Piroue and R.L.Sumner, Phys. Rev. {\bf D11}, 3105(1975).
\bibitem{Gerschel:1988wn}
  C.Gerschel and J.Hufner, Phys. Lett. {\bf B207}, 153(1988).
\bibitem{BraunMunzinger:2000px}
  P.Braun-Munzinger and J.Stachel, Phys. Lett. {\bf B490}, 196(2000).
\bibitem{Thews:2000rj}
  R.L.Thews, M.Schroedter and J.Rafelski, Phys. Rev. {\bf C63}, 054905(2001).
\bibitem{Yao:2017fuc}
  X.Yao and B.M\"uller, Phys. Rev. {\bf C97}, 014908(2018).
\bibitem{Grandchamp:2002wp}
  L.Grandchamp and R.Rapp, Nucl. Phys. {\bf A709}, 415(2002).
\bibitem{Liu:2009nb}
  Y.P.Liu, Z.Qu, N.Xu and P.Zhuang, Phys. Lett. {\bf B678}, 72(2009).
\bibitem{Zhao:2010nk}
  X.Zhao and R.Rapp, Phys. Rev. {\bf C82}, 064905(2010).
\bibitem{Zhou:2014kka}
  K.Zhou, N.Xu, Z.Xu and P.Zhuang, Phys. Rev. {\bf C89}, 054911(2014).
\bibitem{Chen:2015iga}
  B.Chen, Phys. Rev. {\bf C93}, 054905(2016).
\bibitem{Chen:2012gg}
  B.Chen, K.Zhou and P.Zhuang, Phys. Rev. {\bf C86}, 034906(2012).
\bibitem{Zhao:2011cv}
  X.Zhao and R.Rapp, Nucl. Phys. {\bf A859}, 114(2011).
\bibitem{Deng:2012pc}
  W.T.Deng and X.G.Huang, Phys. Rev. {\bf C85}, 044907(2012).
\bibitem{vonWeizsacker:1934nji}
  C.F.von Weizsacker, Z. Phys. {\bf 88}, 612(1934).
\bibitem{Klein:1999qj}
  S.Klein and J.Nystrand, Phys. Rev. {\bf C60}, 014903(1999).
\bibitem{Adam:2015gba}
  J.Adam {\it et al.} [ALICE Collaboration], Phys. Rev. Lett. {\bf 116}, 222301(2016).
\bibitem{Abelev:2012ba}
  B.Abelev {\it et al.} [ALICE Collaboration], Phys. Lett. {\bf B718}, 1273(2013).
\bibitem{Abbas:2013oua}
  E.Abbas {\it et al.} [ALICE Collaboration], Eur. Phys. J. {\bf C73}, 2617(2013).
\bibitem{Bertulani:2005ru}
  C.A.Bertulani, S.R.Klein and J.Nystrand, Ann. Rev. Nucl. Part. Sci. {\bf 55}, 271(2005).
\bibitem{Xie:2015gdj}
  Y.P.Xie and X.Chen, Eur. Phys. J. {\bf C76}, 316(2016).
\bibitem{Yu:2014eoa}
  G.M.Yu, Y.B.Cai, Y.D.Li and J.S.Wang, Phys. Rev. {\bf C95}, 014905(2017).
\bibitem{Yu:2015kva}
  G.M.Yu, Y.C.Yu, Y.D.Li and J.S.Wang, Nucl. Phys. {\bf B917}, 234(2017).
\bibitem{Xie:cite2}
  Y.P.Xie and X.Chen, Nucl. Phys. {\bf A957}, 477(2017).
\bibitem{Shi:2017qep}
  W.Shi, W.Zha and B.Chen,  Phys. Lett. {\bf B777}, 399(2018).

\bibitem{Ma:2013yla} 
  Y.~Q.~Ma, J.~W.~Qiu and H.~Zhang,
  Phys.\ Rev.\ D {\bf 89}, no. 9, 094029 (2014)

\bibitem{Braaten:1996pv} 
  E.~Braaten, S.~Fleming and T.~C.~Yuan,
  Ann.\ Rev.\ Nucl.\ Part.\ Sci.\  {\bf 46}, 197 (1996)

\bibitem{Amundson:1996qr} 
  J.~F.~Amundson, O.~J.~P.~Eboli, E.~M.~Gregores and F.~Halzen,
  Phys.\ Lett.\ B {\bf 390}, 323 (1997)


\bibitem{Bhanot:1979vb}
  G.Bhanot and M.E.Peskin, Nucl. Phys. {\bf B156}, 391(1979).
\bibitem{Zhu:2004nw} 
  X.~l.~Zhu, P.~f.~Zhuang and N.~Xu,
  Phys.\ Lett.\ B {\bf 607}, 107 (2005)


\bibitem{Chen:2016dke}
  B.Chen, T.Guo, Y.Liu and P.Zhuang, Phys. Lett. {\bf B765}, 323(2017).
\bibitem{Fries:2008hs}
  R.J.Fries, V.Greco and P.Sorensen, Ann. Rev. Nucl. Part. Sci. {\bf 58}, 177(2008).
\bibitem{Hwa:2002tu}
  R.C.Hwa and C.B.Yang, Phys. Rev. {\bf C67}, 034902(2003).
\bibitem{Zhao:2017yan}
  J.Zhao and B.Chen, Phys. Lett. {\bf B776}, 17(2018).
\bibitem{ALICE:2012ab}
  B.Abelev {\it et al.} [ALICE Collaboration], JHEP {\bf 1209}, 112(2012).
\bibitem{Shen:2011eg}
  C.Shen, U.Heinz, P.Huovinen and H.Song, Phys. Rev. {\bf C84}, 044903(2011).
\bibitem{Kharzeev:1999bh}
  D.Kharzeev and R.L.Thews, Phys. Rev. {\bf C60}, 041901(1999).
\bibitem{Cooper:1974mv}
  F.Cooper and G.Frye, Phys. Rev. {\bf D10}, 186(1974).
\bibitem{Zha:2017etq}
  W.Zha [STAR Collaboration], J. Phys. Conf. Ser. {\bf 779}, 012039(2017).
\bibitem{G:Baur2002}
  G. Baur, et al., Physics Reports 364 (2002) 359-450
\bibitem{V:Guzey}
  Vadim Guzey's talk in Workshop INT-17-65W, INT, Seattle, (2017)

\bibitem{Vidovic:1992ik}
  M.Vidovic, M.Greiner, C.Best and G.Soff, Phys. Rev. {\bf C47}, 2308(1993).
\bibitem{book:HERA}
J. A. Crittenden, ``Exclusive Production of Neutral Vector Mesons
at the Electron-Proton Collider HERA'' (Springer-Verlag,
Berlin, 1997) 

\bibitem{Khachatryan:2014bva}
  V.Khachatryan {\it et al.} [CMS Collaboration], Phys. Rev. Lett. {\bf 113}, 262301(2014).
\bibitem{Scomparin:2017pno}
  E.Scomparin, Nucl. Phys. {\bf A967}, 208(2017).
\bibitem{Sirunyan:2016znt}
  A.M.Sirunyan {\it et al.} [CMS Collaboration], Phys. Rev. Lett. {\bf 118}, 162301(2017).
\bibitem{Chen:2013wmr}
  B.Chen, Y.Liu, K.Zhou and P.Zhuang, Phys. Lett. {\bf B726}, 725(2013).
\bibitem{Acharya:2017hjh}
  S.Acharya {\it et al.} [ALICE Collaboration], Eur. Phys. J. {\bf C77}, 392(2017).
\bibitem{Adam:2015sia}
  J.Adam {\it et al.} [ALICE Collaboration], Phys. Lett. {\bf B751}, 358(2015).

\end{thebibliography}
\end{document}